\documentclass[letterpaper]{article}

\usepackage[T1]{fontenc}
\usepackage{geometry}
\usepackage{setspace}
\usepackage{achemso}
\usepackage{graphicx}
\usepackage{float}
\newfloat{scheme}{htbp}{los}
\floatname{scheme}{Scheme}
\floatname{chart}{Chart}
\newfloat{graph}{htbp}{loh}
\usepackage{chemformula} 
\usepackage[version = 4]{mhchem} 
\usepackage{authblk}
\usepackage{amsthm}
\usepackage{amssymb}
\usepackage{url}
\usepackage{hyperref}
\usepackage{doi}

\usepackage{mathtools}
\usepackage{physics}
\usepackage{xcolor}
\usepackage{adjustbox}
\usepackage{placeins}
\usepackage{lipsum}
\usepackage{csquotes}
\usepackage{soul}
\usepackage[english]{babel}
\usepackage[utf8]{inputenc}
\usepackage[colorinlistoftodos, color=green!40, prependcaption]{todonotes}
\usepackage{nameref}
\usepackage{dcolumn}
\usepackage{bm}
\usepackage{verbatim}
\setkeys{acs}{doi = true}

\begin{document}

\title{Local Broadband Gigahertz Spectroscopy via SQUID-on-Tip DC Readout}

\author[1]{Senne Vervoort}
\author[1]{Lukas Nulens}
\author[2]{Julia Baumgarten}
\author[4]{Daan B. Boltje}
\author[4]{Jessalyn DeVine}
\author[2]{Alejandro V. Silhanek}
\author[1]{Joris Van de Vondel}
\author[3,4]{Kaveh Lahabi$^*$}
\affil[1]{Quantum Solid-State Physics, Department of Physics and Astronomy, KU Leuven, Celestijnenlaan 200D, Leuven, B-3001, Belgium}
\affil[2]{Experimental Physics of Nanostructured Materials, Q-MAT, Department of Physics, Université de Liège, Allée du 6 Août 19, Sart Tilman B-4000, Belgium.}
\affil[3]{Huygens-Kamerlingh Onnes Laboratory,
Leiden University, 2300 RA Leiden, The Netherlands}
\affil[4]{QuantaMap BV, 2333 CG Leiden, The
Netherlands}

\begin{abstract}
\noindent Superconducting microwave circuits are shaped by their local electromagnetic environment, yet standard microwave measurements do not resolve the local fields and device features that govern their behavior. Here, we combine microwave spectroscopy, nanoscale magnetic sensing, and topographic imaging in a single tapping-mode SQUID-on-tip microscope. The nonlinear Josephson response converts local GHz excitations into Shapiro steps, enabling quantitative measurements of microwave frequency signals and their excitation strength using only DC readout. Applied to a superconducting coplanar waveguide resonator, local spectroscopy reproduces the power-dependent resonance and nonlinear bifurcation observed in microwave transmission measurements. Spatially resolved measurements map the microwave response while simultaneously recording topography and static magnetic features, including vortices and Meissner screening currents. Electromagnetic simulations reveal how coupling through the superconducting probe leads limits spatial resolution and guides localized detector designs. Together, these capabilities establish a platform for correlating microwave circuit dynamics with the local structural and magnetic environment of superconducting devices.
\end{abstract}

\maketitle
*kaveh@QuantaMap.eu
\section{\label{sec:level1} Introduction}
Nanofabricated microwave circuits play a central role in cryogenic quantum technologies and condensed-matter experiments, from circuit quantum electrodynamics in superconducting \cite{Blais2004cQED,wallraff2004strong,van2024advanced} and spin qubits \cite{steinacker2025industry} to sensitive kinetic-inductance detectors for astronomy\cite{Mauskopf2018MKID}. Their operation depends on precise control of on-chip microwave modes and their coupling to mesoscopic elements. Superconducting transmission lines and coplanar waveguide (CPW) resonators provide this control, yet their response is also shaped by local features: vortices, Meissner screening currents, and nanoscale material or fabrication defects can alter current distributions, introduce losses, and create unintended coupling pathways that degrade device performance\cite{gao2008physics,zmuidzinas2012superconducting,nulens2023catastrophic,song2009microwave,Baumgarten2026MagneticLandscape,vervoort2025dc}. In practice, superconducting microwave circuits are primarily characterized through global electrical measurements, which cannot resolve the spatial distribution of microwave fields or establish their connection to mesoscopic features. As quantum circuits become increasingly complex and miniaturized, the lack of spatially resolved characterization tools for microwave driven circuits has become an experimental bottleneck.\\

Addressing this challenge calls for a local cryogenic probe capable of frequency-resolved microwave detection and spatial mapping, while simultaneously resolving the topography and static magnetic landscape of superconducting circuits. Most near-field microwave microscopy techniques operate by measuring the local dielectric and electrical response to microwave excitations, supplied through a scanning tip \cite{barber2022microwave, Vlahacos1996NearField, Berweger2020Nanoelectronic, Geaney2019NearField, Zheng2018Visualization, Anlage2007scanning, wang2025microwave}.  Although well suited to investigating material properties, these techniques provide limited insight into the microwave fields generated by active circuits and lack the required nanoscale magnetic sensitivity. There are also practical constraints: techniques requiring smooth surfaces or fixed-height scanning are poorly suited for the topography of nanofabricated devices, and techniques that expose the sample to external radiation can perturb the operation of cryogenic circuits. Meanwhile, existing NV-center microscopes typically require an externally applied magnetic field, limiting their compatibility with sensitive quantum technologies \cite{Casola2018NV, Borst2023Hybrid, li2026scanningprobequantumsensingmicrowave, ManasValero2026Isofrequency}, while scanning gate microscopy does not provide information on the magnetic landscape\cite{Hegedus2025TLS}.\\

Superconducting quantum interference devices (SQUIDs) offer a promising alternative. As inherently cryogenic sensors with exceptional magnetic sensitivity, SQUID-based microscopy has been widely successful in studying superconducting systems \cite{vlahacos2010cryo,Yang2026ReviewSQUID,sheekey2026squidgraphene, zhang2026squidmeissner} and low-temperature quantum phenomena \cite{Vasyukov2013Magnetometry,Weber2025SQUIDcantilever,zhang2024near, zhang2025ultra, rog2026tapping, roskamp2026wireframe, Halbertal2016thermal, Wyss2022imaging}. Although these studies have largely focused on static signals and low-frequency excitations, the intrinsic dynamics of Josephson junctions extend well into the GHz regime. In practice, exploiting this capability for near-field microwave microscopy presents long-standing challenges, including bandwidth limitations required for direct GHz readout \cite{black1995high, black1995microwave, Talanov2014200MHz}. Previous approaches to circumvent these limitations often require additional microwave signals or complex detection electronics, while others remain restricted to lower frequencies or encounter parasitic resonances at higher frequencies. Additionally, spatial resolution is constrained by large SQUID dimensions or substantial probe-sample separation. Crucially, these implementations do not combine microwave detection with simultaneous topographic imaging, limiting the ability to correlate microwave signals with local device structure.\\

Here, we bring microwave spectroscopy, nanoscale magnetic sensing, and topographic imaging together in a single tapping-mode SQUID-on-tip (SOT) microscope. Our approach exploits the nonlinear Josephson response of the SQUID to convert local microwave excitations into measurable changes in its DC transport characteristics. The resulting Shapiro steps provide a direct measure of the microwave frequency and a means to quantify the coupled microwave power, using only low-frequency electrical readout. We demonstrate this capability on a superconducting CPW resonator, where the locally measured power-dependent resonance closely reproduces those obtained through conventional microwave transmission spectroscopy. We then extend the measurements to spatial imaging, mapping the microwave response across the resonator while simultaneously recording topography and static magnetic signals, including vortices and Meissner screening. We use electromagnetic simulations to analyse our microwave images, and find the achievable spatial resolution is governed by the probe geometry and the coupling mechanism to the SQUID. Together, these findings help guide the development of future SOT architectures capable of simultaneously accessing \textit{in situ} topography, DC magnetic fields, and local microwave fields in superconducting structures and magnetic systems.\\

\section{\label{sec:level2}Measurement principle - SQUID-on-tip}
\begin{figure}[H]
    \centering
    \includegraphics[width=\linewidth]{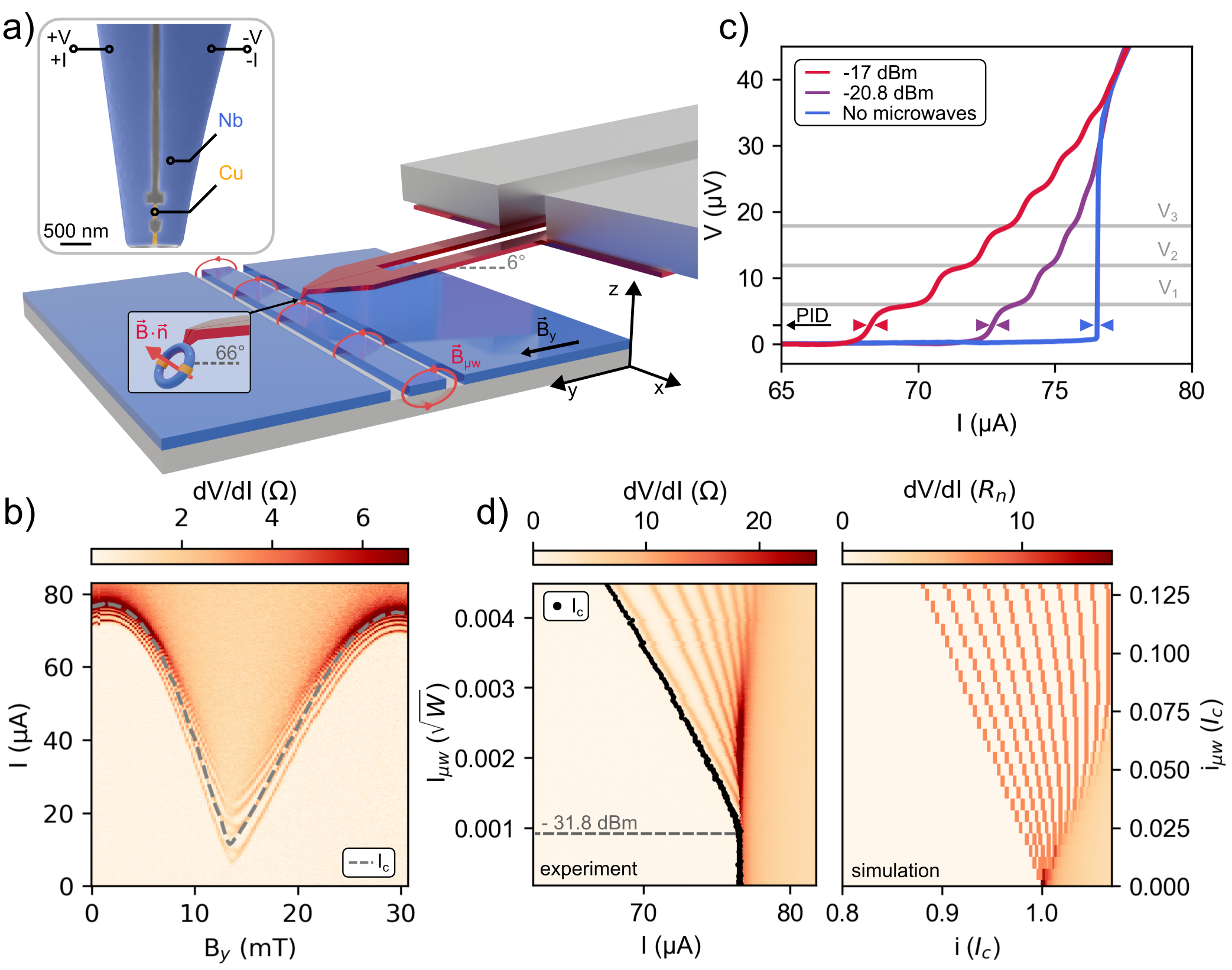}
    \caption{a) Sketch of the QuantaMap SQUID probe showing the spatial orientation of the cantilever with respect to the horizontal sample surface ($xy$-plane). The microwave magnetic ($\vec{B}_{\mu \textrm{w}}$) field in a CPW is indicated by red arrows. The component of the microwave field normal to the SQUID ($\vec{B}\cdot\vec{n}$) is shown in the zoom. $\vec{B}_y$ indicates the applied magnetic field to obtain the data in panel b. The inset shows a false-colored SEM image of the SQUID on the cantilever, where Nb is indicated in blue and Cu in yellow. b) Simultaneous current-field oscillations and Shapiro response with a $-20$ dBm microwave signal applied at 2.873 GHz (color) and the current-field oscillations of $I_c$ with no microwave signal applied (dashed grey line). c) The current-voltage characteristics of the SQUID with no microwave signal applied (blue) and with a microwave signal at 2.873 GHz with a power of $-20.8$ dBm (purple) and $-17$ dBm (red). Expected Shapiro steps are shown with grey lines for $n=1$, 2 and 3. d) Left panel: $dV/dI$ (in color) measured as a function of current and microwave power. The black dots represent the critical current. Right panel: simulation results based on the RSJ model for $dV/dI$ (in color) as a function of current and microwave power.}
    \label{fig:fig1}
\end{figure}

All measurements were performed in a commercial tapping-mode scanning-SQUID microscope of QuantaMap \cite{QuantaMap2026}. It utilizes a DC nano-SQUID, which consists of two proximity Josephson junctions, where the superconducting electrodes (Nb) are coupled through normal metal (Cu) weak links\cite{Clarke2004SQUID}, as shown in the false colored SEM in the inset in Figure \ref{fig:fig1}a. The nano-SQUID is fabricated at the apex of a commercial self-sensing Akiyama atomic force microscopy (AFM) probe, similar to the design in Ref. \citenum{rog2026tapping}. Figure \ref{fig:fig1}a shows a schematic of the AFM probe geometry and the orientation of the SQUID on the cantilever. The tip orientation with respect to the microwave fields in a CPW is indicated, where the tip is tilted by 66° with respect to the sample surface, while the probe arms are inclined by 6°. The SQUID response to an applied magnetic field is determined by the flux threading the SQUID loop \cite{Clarke2004SQUID} and therefore, only the normal component of the magnetic field, $\vec{B}\cdot \vec{n}  = \sin(66^\circ)\,B_y + \cos(66^\circ)\,B_z$, contributes to the SQUID response. To establish the DC operating characteristics of the SQUID, we first characterize its magnetic field dependence. The grey dashed line in Figure \ref{fig:fig1}b shows typical SQUID oscillations of the critical current as function of applied magnetic field in the $y$-direction, indicated as $\vec{B}_y$ in Figure \ref{fig:fig1}a. The oscillations have a period of $\Phi_0/A$, with $\Phi_0$ the flux quantum and $A$ the area of the SQUID \cite{Jaklevic1964SQUID,Clarke2004SQUID}. The measured oscillation period, $B_{\textrm{osc}} = 28.3$ mT, corresponds to a SQUID area of $A = \frac{\Phi_0}{B_{\textrm{osc}}\sin{66^\circ}} = 0.08$ $\mu\text{m}^2$, in good agreement with the geometrical area of the device as obtained from SEM imaging ($A= 0.06$ $\mu\text{m}^2$)  \cite{Ketchen1985SQUIDarea}. The current-voltage characteristic at an applied magnetic field of 0 mT is represented by the blue curve in Figure \ref{fig:fig1}c, where the SQUID reaches a maximum critical current $I_c =$ 76.5 $\mu$A. The normal-state resistance of the SQUID is $R_n = 2.6$ $\Omega$, resulting in a characteristic voltage $V_c = R_n I_c = 199$ $\mu$V. \\

When the junction is driven by a microwave field of frequency $f_{ac}$, constant-voltage steps appear at integer multiples of $\Phi_0 f_{ac}$ \cite{Josephson1962,Shapiro1963steps}, directly linking the drive frequency to quantized voltages and enabling metrological applications \cite{Hamilton1989voltstandard,kohlmann2007josephson}. 
In the current‑biased case, the nonlinear junction dynamics are governed by the intrinsic Josephson timescale set by the characteristic frequency $f_c = V_c/\Phi_0 = 96.2$ GHz, which defines the natural dynamical response of the device. Shapiro steps occur most robustly when the applied microwave frequency is below or comparable to this intrinsic scale, $f_{ac} \lesssim f_c$. For $f_{ac} > f_c$, the junction cannot follow the drive on its natural timescale, and the steps are suppressed \cite{Likharev1986JJ}. As such, the device is ideally suited for probing the $1$--$20$ GHz frequency range targeted in many of the aforementioned applications.\\

As a proof of principle for microwave detection, we position the SQUID, with temperature 2.9 K, in tapping mode at 300 nm from a CPW transmission line at 870 mK carrying a microwave signal at $f_{ac} = 2.873$ GHz. Details on the microwave characterization of the sample can be found in SI \ref{sec:sup_freq_SQUID}.\\

Figure \ref{fig:fig1}b shows $dV/dI$ as function of current and magnetic field when driving the waveguide with a $-20$ dBm microwave signal, which corresponds to $B_{\mu{w}} = 1.8$ $\mu$T (all reported powers represent the power arriving at the waveguide, see SI \ref{sup:Bsamp}).
Compared to the gray dashed line (no microwave), it demonstrates a reduction of the critical current and the appearance of Shapiro steps. The presence of the steps becomes more apparent in the purple and red current-voltage characteristics at 0 mT in Figure \ref{fig:fig1}c for excitation powers of $-20.8$ dBm (1.6 $\mu$T) and $-17$ dBm (2.5 $\mu$T), respectively. Gray lines in Figure \ref{fig:fig1}c corresponding to $V_n = n\Phi_0f_{ac}$, with $n$ an integer, show perfect correspondence between the expected step height and the experimental value.
It is important to understand the coupling mechanism between the SOT and the microwave magnetic field, as the coupling will drastically impact the interpretation of the imaging results presented in sections \ref{sec:level4} and \ref{sec:level5}. Solutions of the RSJ model (see SI \ref{sup:even}) show that only even Shapiro steps ($V_{2n}$) would be present if the coupling were determined by a microwave \textit{field} threading the SQUID. In contrast, the curves in Figure \ref{fig:fig1}c exhibit both even and odd steps, as would be the case for a SQUID driven by a microwave \textit{current}. We propose that the superconducting leads contacting the SQUID act as antennas and convert the incident microwave field into an AC current through the SQUID, analogous to the mechanism underlying AC susceptibility measurements \cite{ACsus}. Consequently, the SQUID is expected to be sensitive not only to the incident microwave field through the SQUID but also to the microwave currents induced in the superconducting leads, an interpretation that is further supported by the spatial-resolution analysis presented in Section \ref{sec:level5}.

The microwave conversion introduces an \textit{a priori} unknown coupling factor between the field amplitude in the waveguide and the current amplitude in the SQUID. Because this process depends on the SQUID geometry and the spatial electromagnetic field distribution, the conversion factor is difficult to estimate accurately. To extract the effective microwave current through the SQUID, we compare the measured power dependence of the Shapiro steps with theoretical predictions derived from the RSJ model in Figure \ref{fig:fig1}d. Figure \ref{fig:fig1}d (left panel) shows $dV/dI$ as a function of $I$ and the square root of microwave drive power to express power in units of current, with the critical current indicated by black dots. A measurable suppression of $I_c$ is first observed at an input power of $-31.8$ dBm (0.5 $\mu$T), indicated by the gray dashed line in the figure. Above this power, $I_c$ decreases and Shapiro steps become visible, as illustrated by the purple and red current-voltage characteristics in Figure \ref{fig:fig1}c.
Above the detection threshold, $I_c$ exhibits a linear dependence on the applied microwave current. This behavior is consistent with the RSJ simulations of a microwave current driven junction shown in Figure \ref{fig:fig1}d (right panel), where the normalized DC current $i$ and microwave drive amplitude $i_{\mu \textrm{w}}$ are expressed in units of $I_c$ and $dV/dI$ in units of $R_n$. The simulations reproduce the observed linear suppression of the critical current with increasing drive amplitude. 
However, a discrepancy is observed between simulations and experiments: the decrease in $I_c$ occurs only at higher $I_{\mu\textrm{w}}$ in the measurements. 
Below the detection limit of $-31.8$ dBm, the critical current is determined by thermal fluctuations, not included in the simulations \cite{massarotti2015switch,baumans2017statistics,dausy2021switch}. Only when the microwave drive exceeds this thermal noise floor, $I_c$ decreases as function of $I_{\mu\textrm{w}}$.
By comparing the reduction in $i_c$ in the simulations with the reduction in the experimentally obtained $I_c$, we can estimate $I_{\mu\textrm{w}}$ flowing through the SQUID.\\

For spatially resolved measurements, $I_c$ must be determined rapidly to keep the overall measurement time reasonable. To this end, a PID control loop is used to dynamically adjust the bias current such that the voltage across the junction is maintained at half of the first Shapiro step, $V_1/2$ (arrow and triangles in Figure \ref{fig:fig1}c). The resulting feedback-adjusted current is recorded and serves as an effective measure of the critical current. In this scheme, the junction voltage acts as the input variable for the control loop. In addition, a small AC current excitation is superimposed on the bias current to enable simultaneous lock-in detection of the differential voltage, providing access to the differential resistance. 
Further details on the measurement scheme are presented in SI \ref{Sup:setup_Leiden}. By scanning the probe across the structure and recording the changes in $I_c$, the spatial profile of the microwave field can be reconstructed. Repeating this procedure for different frequencies enables the identification of resonances and their spatial mode structure.\\

\section{\label{sec:level3}Broadband Microwave Spectroscopy with SQUID-on-tip}
To demonstrate the imaging capacity of the SOT probe, a hanger-type $\lambda/4$ CPW resonator was fabricated. Such CPW structures are widely used in superconducting \cite{Nulens2025,kalashnikov2024demonstration} and quantum technologies \cite{wallraff2004strong}, making them a relevant and realistic test platform. In CPW geometries, both electric and magnetic microwave fields are spatially confined\cite{pozar2011microwave}, and thereby well suited for quantifying the spatial resolution of the measurement technique. Furthermore, in a $\lambda/4$ resonator the magnetic and electric field distributions are spatially separated; the magnetic field is maximum at the shorted end and vanishes at the open end, while the electric field shows the opposite behavior. This enables distinguishing the sensitivity of the probe to electric and magnetic components of the microwave field. At the shorted end of the resonator the current is enhanced with respect to the feedline by a factor of $\sim 1.6\sqrt{Q_l}$ \cite{pozar2011microwave}, where $Q_l$ is the loaded quality factor of the resonator. Since the microwave magnetic field scales with current density, this results in a significantly enhanced magnetic field at the resonator end compared to the feedline, leading to a stronger local microwave signal. These properties provide a controlled platform to investigate the spatial SOT response and its sensitivity to electric and magnetic fields.\\

\begin{figure}[H]
    \centering
    \includegraphics[width=\linewidth]{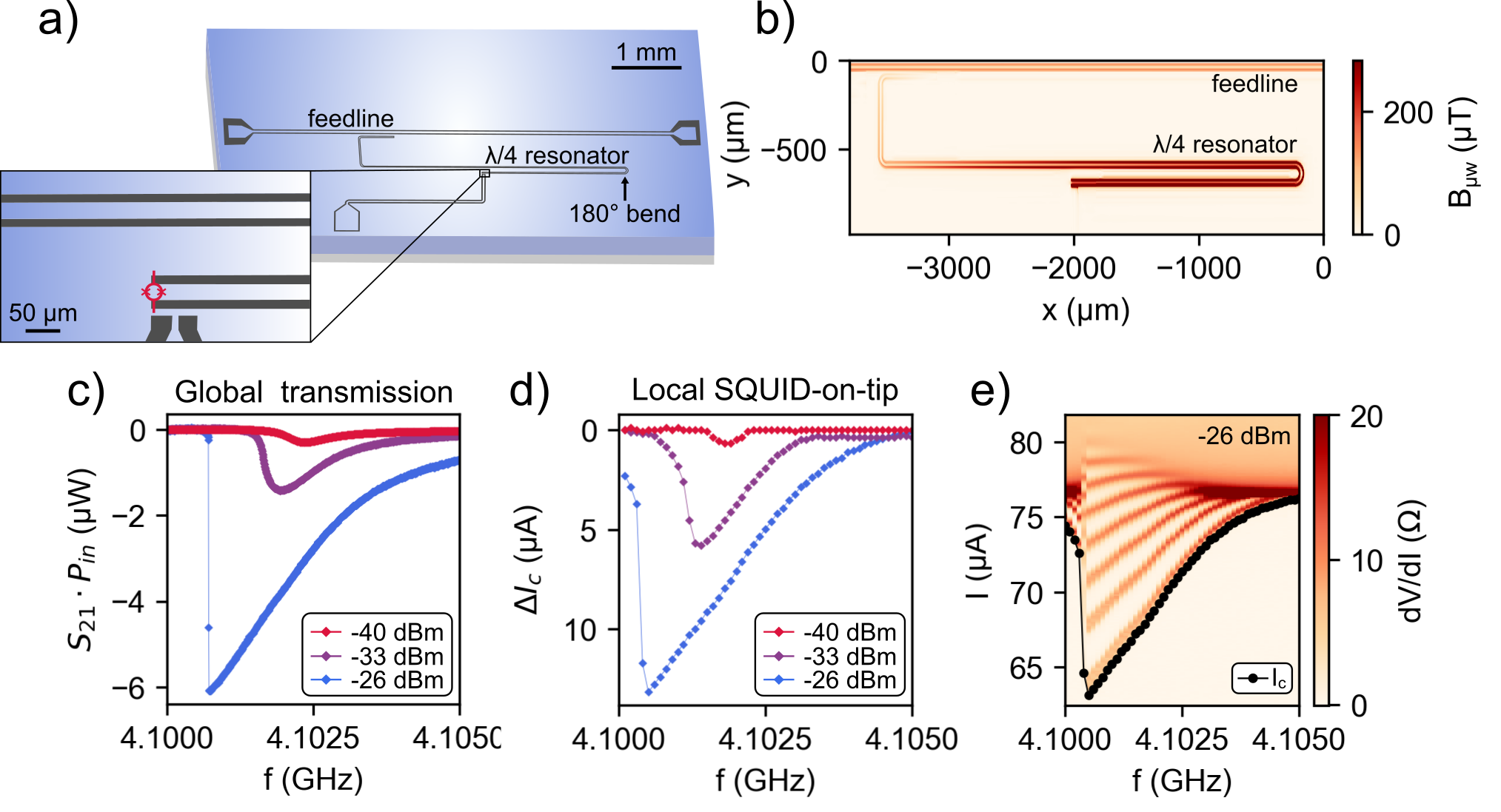}
    \caption{a) Layout of the $\lambda/4$  CPW resonator. The inset shows the shorted end of the resonator with the SQUID position in red. b) Variation of the microwave magnetic field amplitude through the SQUID as a function of position along the $\lambda/4$ resonator, extracted from Sonnet simulations with 13 dBm driving power. c) $S_{21}$ of the CPW resonator scaled by the input power as function of frequency from the linear (red curve, $-40$ dBm) to the non-linear regime (purple and blue curve $-33$ dBm and $-26$ dBm, respectively). d) Change in the critical current of the SQUID as a function of frequency for the same input powers as in panel c. e) $dV/dI$ as function of current and frequency driven at $-26$ dBm. Black dots indicate the critical current and correspond to the blue data points in panel d.}
    \label{fig:fig2}
\end{figure}

Figure \ref{fig:fig2}a shows the design of the 50 nm-thick Nb CPW structure, including the central feedline to which a single $\lambda/4$ resonator of length 6.18 mm is capacitively coupled. The inset details the shorted end of the resonator, with a flux line to possibly bias the resonator with a magnetic field, similar as in Ref. \citenum{Nulens2025}. The CPW dimensions are designed to realize a 50 $\Omega$ impedance-matched transmission line, and the structure is fabricated on a high-resistivity Si substrate. Fabrication details are provided in SI \ref{Sup:fab}. 
For this device geometry, method-of-moments simulations in Sonnet \cite{Sonnet} were performed. Figure \ref{fig:fig2}b shows the microwave magnetic field amplitude through the SQUID, extracted from the Sonnet simulations, along the resonator structure. These simulations will serve as a basis for further analysis.
The sample was measured in a Bluefors cryostat equipped with the scanning SOT, and global transmission measurements were performed using a VNA. Details of the experimental setup can be found in the SI \ref{Sup:setup_Leiden}. To distinguish reductions in $I_c$ arising from microwave irradiation from those induced by static magnetic fields, all microwave measurements were performed at $B_y = 0$ mT. Figure \ref{fig:fig2}c shows the change in transmission ($S_{21}$), scaled by the input power, at a base temperature of 870 mK, obtained via conventional transmission measurements. The sample shows a resonance frequency of $f_r = 4.1024$ GHz (red curve) and a loaded quality factor of 5030. The response of the resonator depends strongly on the applied microwave power, as demonstrated by conventional transmission measurements. As the drive power is increased from $-40$ dBm to $-26$ dBm, the resonance shifts towards lower frequencies due to the onset of non-linear behavior in the resonator. This non-linearity leads to a bifurcation of the resonance curve, resulting in a pronounced jump on the low-frequency side at higher powers (purple and blue traces)\cite{swenson2013operation}.\\

In addition to global transmission measurements, the resonator can also be characterized using the SOT as a local probe. Since the microwave magnetic field is enhanced at the shorted end of the resonator due to the increased current density (Figure \ref{fig:fig2}b), the SQUID is positioned in this region to maximize its response to the microwave signal, as indicated by the red symbol in the inset of Figure \ref{fig:fig2}a. 
Since there is an enhanced microwave field for a given drive power, the expected detection limit is reduced. Nevertheless, the linear dependence of $I_c$ on microwave power (Figure \ref{fig:fig1}d) remains valid as long as the microwave-induced modulation exceeds thermal fluctuations.
This enables local microwave spectroscopy using the SQUID as a sensitive power detector.
Figure \ref{fig:fig2}e shows the recorded $dV/dI$ as function of current through the SQUID and frequency for an applied microwave power of $-26$ dBm. A reduction in $I_c$ (black dots) is observed only at the resonance frequency of the resonator due to the enhanced microwave fields at resonance. Furthermore, Shapiro steps can be seen up to $n=7$. 
Figure \ref{fig:fig2}d shows the the change in critical current $\Delta I_c$ as function of frequency for different drive powers. Upon increasing the microwave drive power, the same bifurcation behavior observed in the transmission measurements in Figure \ref{fig:fig2}c is also reflected in the DC response of the SQUID in Figure \ref{fig:fig2}d. The agreement between $S_{21}$ measurements and the SQUID critical current demonstrates that the SOT can serve as a local broadband microwave probe.\\

\section{\label{sec:level4} Spatially resolved probing of a CPW superconducting resonator with SQUID-on-tip}
\begin{figure}[H]
    \centering
    \includegraphics[width=0.9\linewidth]{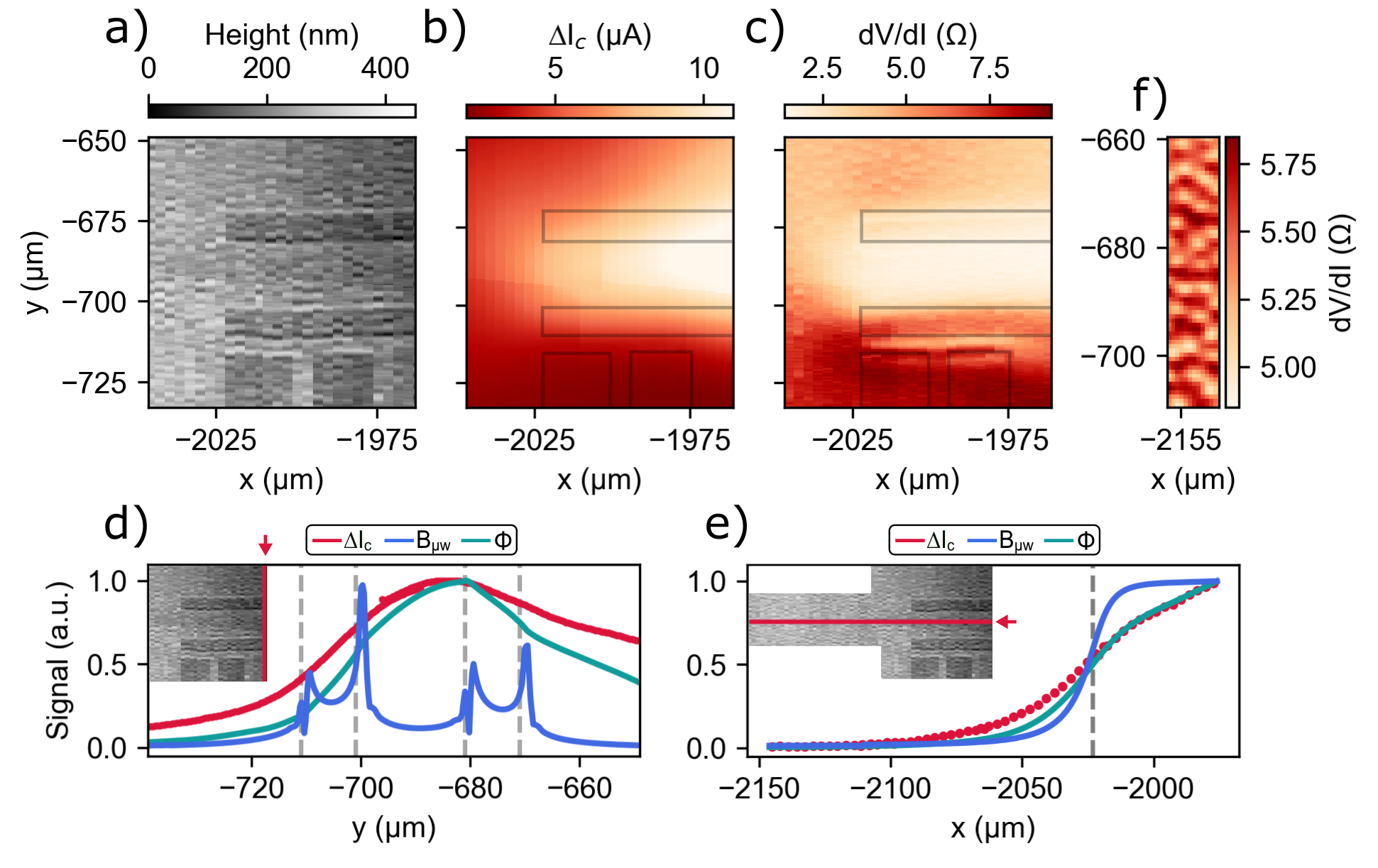}
    \caption{a) Topographic image of the shorted end of the resonator. b) $\Delta I_c$ as function of $x$ and $y$ coordinate, simultaneously obtained with the topography image. Grey lines correspond to the edges of the resonator. c) $dV/dI$ as function of $x$ and $y$ coordinate, simultaneously obtained with the topography and $\Delta I_c$, respectively. d) Normalized change in critical current (red), simulated amplitude of the microwave magnetic field (blue), and simulated total flux through the probe (green) along the red vertical line in the inset. The gray dashed lines represent the edges of the resonator. e) Normalized change in critical current (red), simulated amplitude of the microwave magnetic field (blue) and simulated total flux through the probe (green) along the red horizontal line in the inset. The gray dashed line represents the shorted end of the resonator. f) Higher resolution $dV/dI$ as function of position, permitting to resolve individual vortices around $x=-2155$ $\mu$m.}
    \label{fig:fig3}
\end{figure}

In the previous section, we demonstrated that the SOT is sensitive to the magnetic component of an applied microwave field, and that its critical current can serve as a probe for the field amplitude. We now extend this approach to spatially resolve microwave fields as a function of position at resonance. Figure \ref{fig:fig3}a shows the topography of the device, determined by the AFM capabilities of the SOT. Figures \ref{fig:fig3}b and c show the simultaneously acquired spatial maps of $\Delta I_c$ and $dV/dI$, respectively. For clarity, the resonator geometry is overlaid as black contours in Figures \ref{fig:fig3}b and c. All measurements were performed using the PID-based feedback scheme described in Section \ref{sec:level2}, with an input power of $-29$ dBm.\\

In Figure \ref{fig:fig3}b, a pronounced increase in $\Delta I_c$ is observed above the shorted end of the resonator (white region), where the microwave field is expected to reach its maximum amplitude. Away from the resonator and into the ground plane, $\Delta I_c$  recovers to zero (red region).
The minimum measured critical current is 64.2 $\mu$A and, based on RSJ simulations at the resonant frequency and following the approach used in Figure \ref{fig:fig1}d, we find that a reduction in $I_c$ of 12.3 $\mu$A  corresponds to a maximum detected microwave current amplitude of 13.9 $\mu$A in the SQUID.
Although the reduction in $I_c$ is strongest near the resonator, Figure \ref{fig:fig3}b reveals a clear asymmetry, manifested as a tail extending toward the upper right corner. This feature becomes more apparent in Figures \ref{fig:fig3}d and e, which show line scans of the field as a function of the $y$ and $x$ coordinates, respectively.\\

The line scans are taken along the red lines indicated in the insets with an arrow, with the dashed gray lines marking the resonator edges. Red dots show the normalized $\Delta I_c$, while the blue lines show the normalized magnetic field through the SQUID, extracted from Sonnet simulations. The simulated magnetic field is strongly concentrated at the CPW edges and decays rapidly within the ground plane, consistent with the expected edge-dominated current distribution in CPW structures \cite{pozar2011microwave,li2026scanningprobequantumsensingmicrowave}. 
However, in the experiment, these sharp features are washed out, and a non-zero change in critical current remains present even when the SQUID has moved beyond the edge of the resonator. These observations are attributed to the point spread function (PSF) of the probe. In Section \ref{sec:level2}, we have shown that a microwave current drives the SQUID, rather than a microwave field through the SQUID, rendering the measured response sensitive to the convolution of the CPW field profile with the spatial distribution of the leads. Therefore, it is the contact lead geometry which determines the PSF, not the geometry of the SQUID, as is the case for DC field measurements \cite{ullmann2026pointspread}.
Furthermore, Sonnet simulations show that the electromagnetic fields extend up to 80 $\mu$m above the waveguide (see SI \ref{sup:height}), below which the contact leads on the sloped tip can still feel a microwave flux.\\

To verify that the microwave current responsible for the Shapiro steps originates from flux through the superconducting contact leads, we use the results from the Sonnet simulations to calculate the total flux through the extended probe geometry. Details can be found in SI \ref{sup:sonnet}. The resulting normalized total flux through the probe, shown as green curves in Figures \ref{fig:fig3}d and e, reproduces the main experimental trends. Specifically, in Figure \ref{fig:fig3}d the flux peaks above the resonator, decays gradually into the ground plane to a non-zero value, and exhibits a kink at the point where the SQUID crosses the resonator edge. Likewise, the model incorporating the flux through the probe also captures the main features of the horizontal line scan shown in Figure \ref{fig:fig3}e.

In addition to its microwave sensitivity, the SQUID remains sensitive to DC magnetic fields. The simultaneously acquired differential resistance (Figure \ref{fig:fig3}c) reflects the position-dependent changes in the slope of the current voltage characteristics. Since the differential resistance at half the first Shapiro step depends sensitively on the magnetic flux through the SQUID (see Figure \ref{fig:fig1}b), it allows detection of small DC magnetic features even when the $I_c$ reduction is dominated by microwave excitation. In Figure \ref{fig:fig3}c, inside the T-piece a red (and white) part is present, representing an increased (decreased) differential resistance. These are Meissner currents flowing along the edge of the resonator, with the color contrast reflecting their opposite polarity. Despite the limited pixel density of the images, vortices can still be discerned as dots in the upper part of the panel. To image these vortices without contributions from other static features near the resonator end, the SQUID was positioned above the ground plane ($x$ = -2155 $\mu$m) and an image was made with a reduced scan speed. The resulting image is shown in Figure \ref{fig:fig3}f, where each red and white feature corresponds to an individual vortex. \\

\section{\label{sec:level5} Effect of point spread function on spatial resolution}
\begin{figure}[H]
    \centering
    \includegraphics[width=0.9\linewidth]{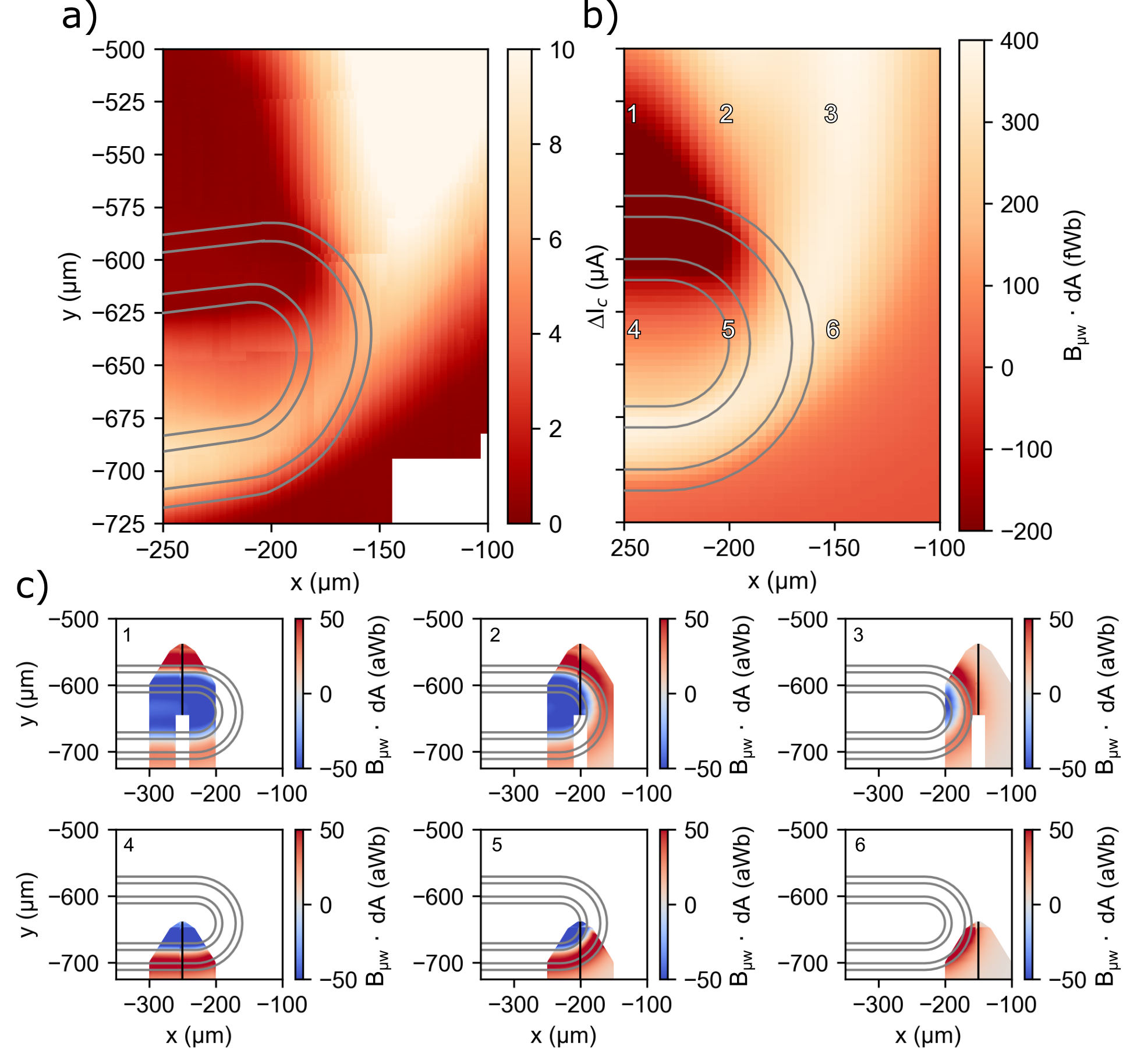}
    \caption{a) Measured $\Delta I_c$ as a function of position near the 180$^\circ$ bend. The resonator edges are overlaid in grey. b) Simulated magnetic flux integrated over the complete probe geometry as a function of position near the 180$^\circ$ bend of the resonator. Simulations were performed with 13 dBm input power. c) Simulated spatial distributions of magnetic flux through the probe geometry for the six positions indicated in panel b.}
    \label{fig:fig4}
\end{figure}

To further investigate the effect of the probe specific PSF, we measure $\Delta I_c$ as a function of position at the 180$^\circ$ turn of the resonator, indicated in Figure \ref{fig:fig2}a. Figure \ref{fig:fig4}a shows the acquired $\Delta I_c$ at this position, with the topography as extracted from the simultaneously acquired AFM overlaid in gray for clarity. The image is skewed due to a piezo non-linearity. Here, the white region of high detected power extends far up to the upper right corner, and seems to become completely unrelated to the topography. To verify that this effect is caused by the coupling of the magnetic field to the superconducting contact leads, we calculate the total flux through the probe geometry as function of position, shown in Figure \ref{fig:fig4}b. Indeed, the simulations reproduce the smearing out into the upper right corner that is seen in the experimental data, confirming that the coupling is largely mediated by the superconducting contact leads. Figure \ref{fig:fig4}c presents the flux distributions within the contact geometry at the six positions indicated in panel b, where red represents positive flux through the probe, while blue represents negative flux. In positions 1-3, while the SQUID is positioned at the tip of the probe, a magnetic field is still felt by the probe even when the SQUID is not above the resonator. The detection is more local when less area of the contact leads is above the resonator (positions 4-6). 
However, a notable discrepancy occurs near position 1, where the simulations predict a finite negative flux signal while the measured critical current remains essentially unchanged. This inconsistency arises because the model considers only the total flux coupled to the probe, thereby discarding information about the phase of the flux in the left and right arms of the pickup loop.
The relative sign of the flux in each arm governs the direction of the induced microwave currents within the probe geometry and, consequently, the microwave current flowing through the SQUID.
As shown in panel c for position 1, a large and nearly symmetric region of the contact geometry overlaps with the resonator, resulting in flux distributions that are approximately symmetric with respect to the center of the probe. Under these conditions, oppositely directed microwave currents induced in different parts of the probe can cancel, yielding a negligible net microwave current through the SQUID despite the presence of substantial magnetic flux. More results from these simulations at different positions can be found in SI \ref{sup:bend} . While the model successfully captures the dominant experimental trends, it does not capture effects arising from the spatial distribution of the polarity of the flux coupled to individual probe arms.

\section{Conclusion and outlook}
In conclusion, we demonstrate a SQUID-on-tip platform that combines nanoscale sensitivity to static magnetic fields with microwave spectroscopy through its Shapiro response, where the current detection sensitivity is limited primarily by the probe temperature. Measurements of a coplanar waveguide resonator show close agreement between SQUID-based spectroscopy and conventional microwave transmission measurements, including nonlinear resonator dynamics, while spatial scanning establishes the capability for microwave imaging. Experimentally validated electromagnetic simulations reveal that coupling through the superconducting contact leads dominates the point spread function and currently limits spatial resolution. This insight provides a rational design strategy for next-generation probes, where suppressing nonlocal pickup could substantially improve imaging fidelity. Looking ahead, the demonstrated technique could enable local spectroscopy of individual resonators in densely integrated superconducting circuits, from microwave kinetic inductance detector arrays \cite{rasvinke2026singlephoton,Reyes2026KIDarray} to quantum processors \cite{hertzberg2021laseranneal}. By correlating their microwave response with local magnetic and structural features, this approach could help identify frequency collisions, parasitic coupling, and fabrication defects. More broadly, combining microwave spectroscopy, nanoscale magnetic sensing, and topographic imaging in a single scanning platform opens new opportunities for investigating microwave-driven phenomena in quantum materials, superconducting circuits, and chip-based qubits with a single scanning platform.

\section*{Acknowledgments}
S.V. acknowledges support from FWO research fellowship nr. 11A5426N and under the grant Weave G0D7723N. The authors acknowledge support from the EUCOST action SUPERQUMAP (CA21144) and the Fonds de la Recherche Scientifique-FNRS under the grant Weave - PDR T.0208.23 and CDR J.0199.25. 
This research is supported and funded by an interuniversity BOF project (IBOF-23-065). This work was financed by the Dutch National Growth Fund, as part of the Quantum Delta NL Programme in Project SME\_R027\_MQC.
This work was carried out during a sabbatical stay of A.V.S at KU Leuven, funded by a scientific mission of the Fonds de la Recherche Scientifique – F.R.S.-FNRS (Belgium). A.V.S. thanks the University of Liège for granting this research leave, and KU Leuven for its hospitality.

\section*{Author contributions}
L.N designed and fabricated the CPW devices. D.B., J.D., S.V., and L.N. executed the SQUID-on-tip and microwave measurements presented in this manuscript. J.B. performed the simulations, supported by A.V.S.
The manuscript was written by S.V., L.N., and J.V.V. with the help of all other authors. All authors discussed the results and reviewed the manuscript. K.L. and J.V.V. initiated the research.
\section{Supplementary information}
\subsection{Fabrication details}
\label{Sup:fab}
For the fabrication of the niobium CPW resonator, the resonator pattern was first defined by electron-beam lithography (EBL) on a 525-\textmu m-thick high-resistivity silicon substrate ($\rho > 20~\mathrm{k}\Omega\cdot\mathrm{cm}$). A 4-nm-thick titanium sticking layer and a 50-nm-thick niobium layer were then deposited by MBE at room temperature, at a pressure below $1\times10^{-9}$ mbar, with deposition rates of $0.2~\text{\AA}/$s and $1.2~\text{\AA}/$s for Ti and Nb, respectively. The resonator was subsequently defined using a standard lift-off process. The $\lambda/4$ resonator has a length of 6.18 mm and was designed to have a resonance frequency of 4.8 GHz.

\subsection{Experimental setup}
\label{Sup:setup_Leiden}
The tapping-mode SOT setup used for this work is similar to that described by Rog \textit{et al.} \cite{rog2026tapping}. A SQUID is fabricated at the apex of a commercial AFM probe. This probe is mounted in a closed-cycle helium refrigerator along with the sample, both attached to separate piezo stacks for coarse (sample) and fine (probe) movement. By recording the SQUID voltages while simultaneously performing tapping-mode AFM, the various quantities detectable by the SQUID can be mapped to topographical features on the sample.\\

\begin{figure}[H]
    \centering
    \includegraphics{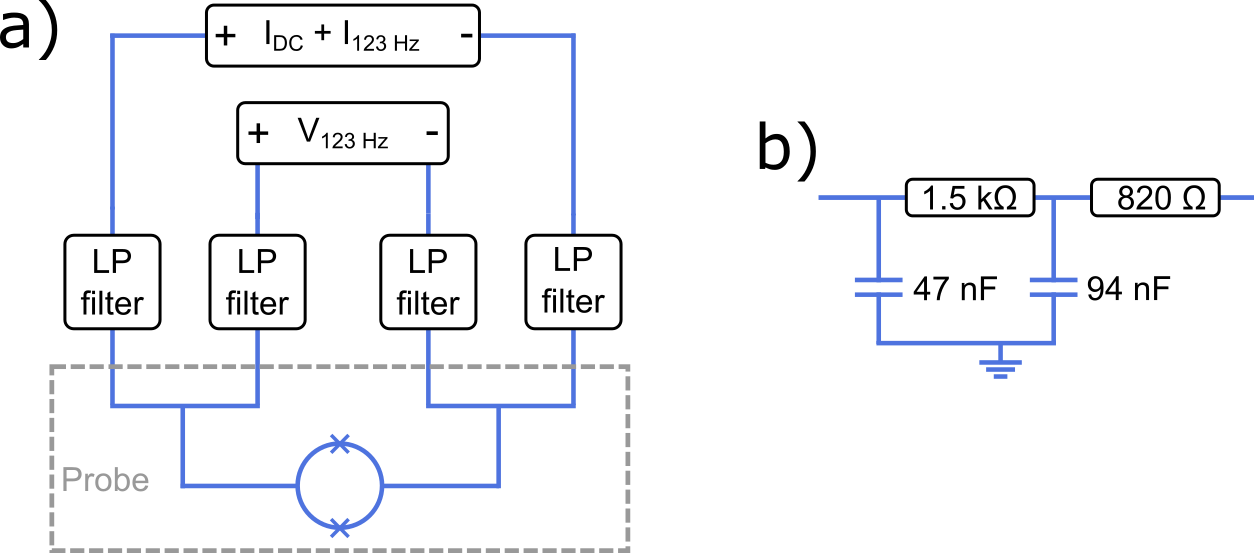}
    \caption{a) Electrical connections to the probe containing the SOT. All lines are filtered using in-house made low-pass (LP) filters with a cut-off frequency around 1 kHz. b) Schematic of the in-house made low-pass filter used to filter the electrical connections to the probe.}
    \label{fig:sup_fig_5}
\end{figure}

Figure \ref{fig:sup_fig_5}a shows the wiring configuration used to measure the SQUID response to microwave fields for this work. All SQUID lines were filtered using low-pass ladder filters with cut-off frequency around 1 kHz. The schematic of the low-pass filters is shown in Figure~\ref{fig:sup_fig_5}b, with the corresponding resistance and capacitance values indicated.\\

\begin{figure}[H]
    \centering
    \includegraphics{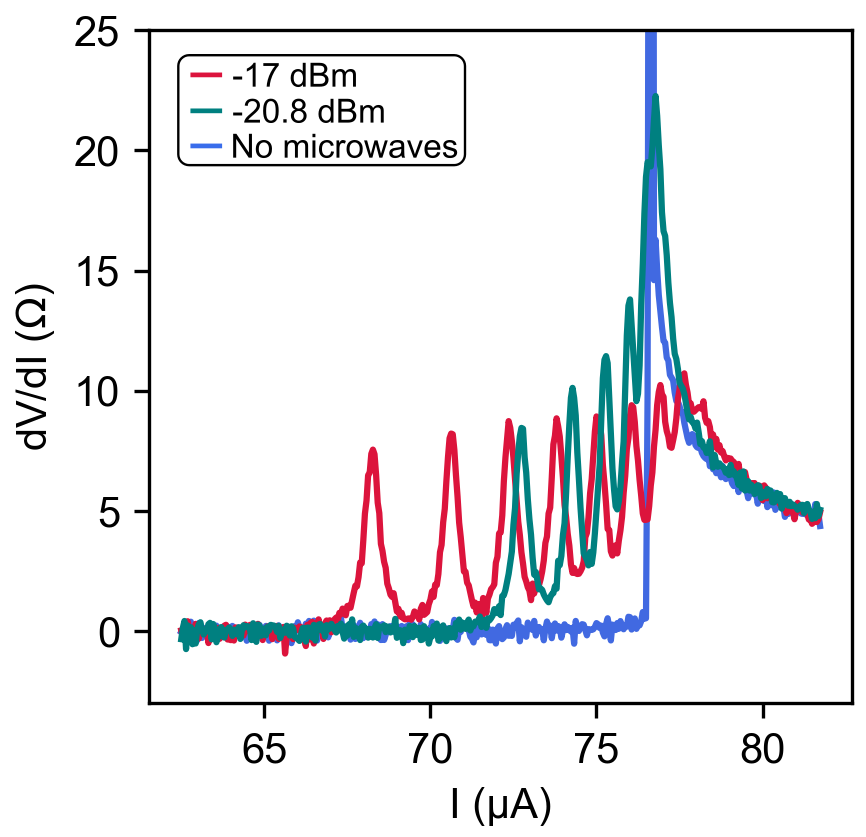}
    \caption{Measured $dV/dI$ curves used to obtain the $IV$ curves shown in Figure \ref{fig:fig1}c. The curves correspond to no applied microwave power through the feedline (blue), $-20.8$ dBm (green), and $-17$ dBm (red).}
    \label{fig:sup_fig_3}
\end{figure}

The SQUID was current-biased using a multi-frequency lock-in amplifier (MFLI; Zurich Instruments) in series with a resistance of 10.4 k$\Omega$. A small AC current with amplitude $dI = 48$ nA and frequency 123 Hz was supplied on top of the DC current. The resulting differential voltage $dV$ at 123 Hz was measured by the MFLI. The differential resistance $dV/dI$ as function of current was obtained at different microwave powers with the tip positioned above the CPW transmission line. The $IV$ curves shown in Figure \ref{fig:fig1}c were obtained by integrating the corresponding $dV/dI$ curves in Figure \ref{fig:sup_fig_3}.

The input and output of the central feedline on the resonator sample were connected to two semi-rigid coaxial cables. CW-RF signals (Signal Core; SC5511A) were passed through three $-6$ dBm attenuators followed by a bandpass filter (Quantum Microwave; QMC-CRYOBPF-3309) prior to entering the sample feed line. The RF output traversed through a bandpass filter (Quantum Microwave; QMC-CRYOBPF-3309) and then was detected using a diode to monitor the relative transmission through the entire setup.

\subsection{Frequency dependent coupling between the microwave and SQUID probe.}
\label{sec:sup_freq_SQUID}
\begin{figure}[H]
    \centering
    \includegraphics{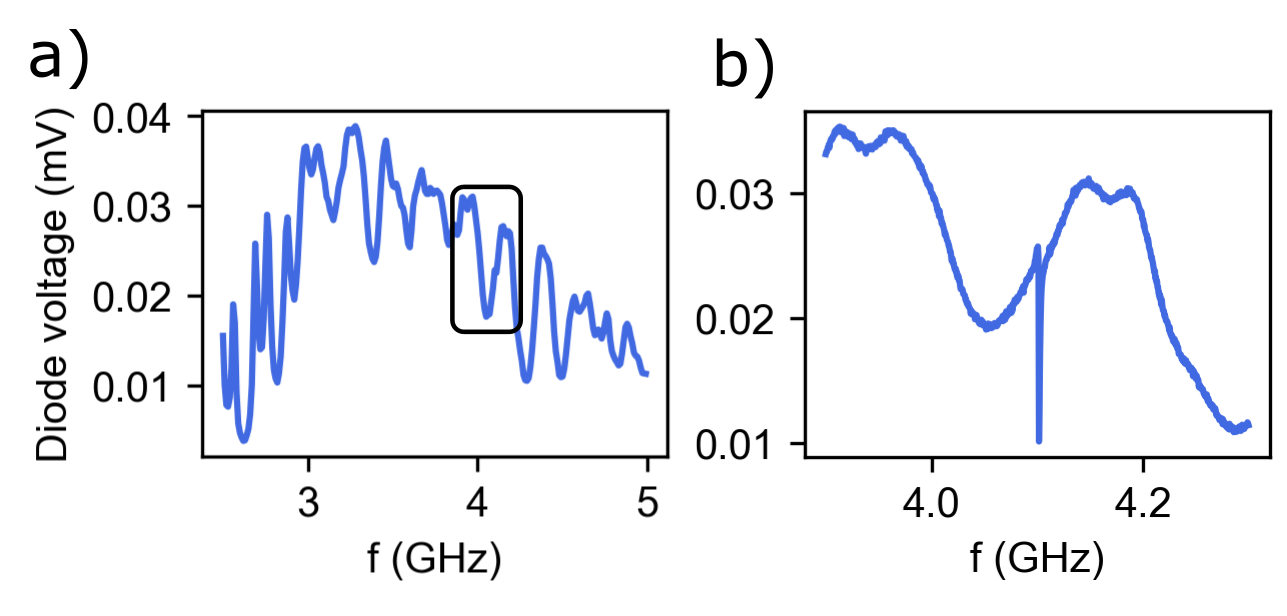}
    \caption{a) Diode voltage as function of frequency, used to estimate the transmission parameter. Large oscillations are present on top of the resonance. b) Zoom in on the resonance feature.}
    \label{fig:diode}
\end{figure}

\begin{figure}[H]
    \centering
    \includegraphics{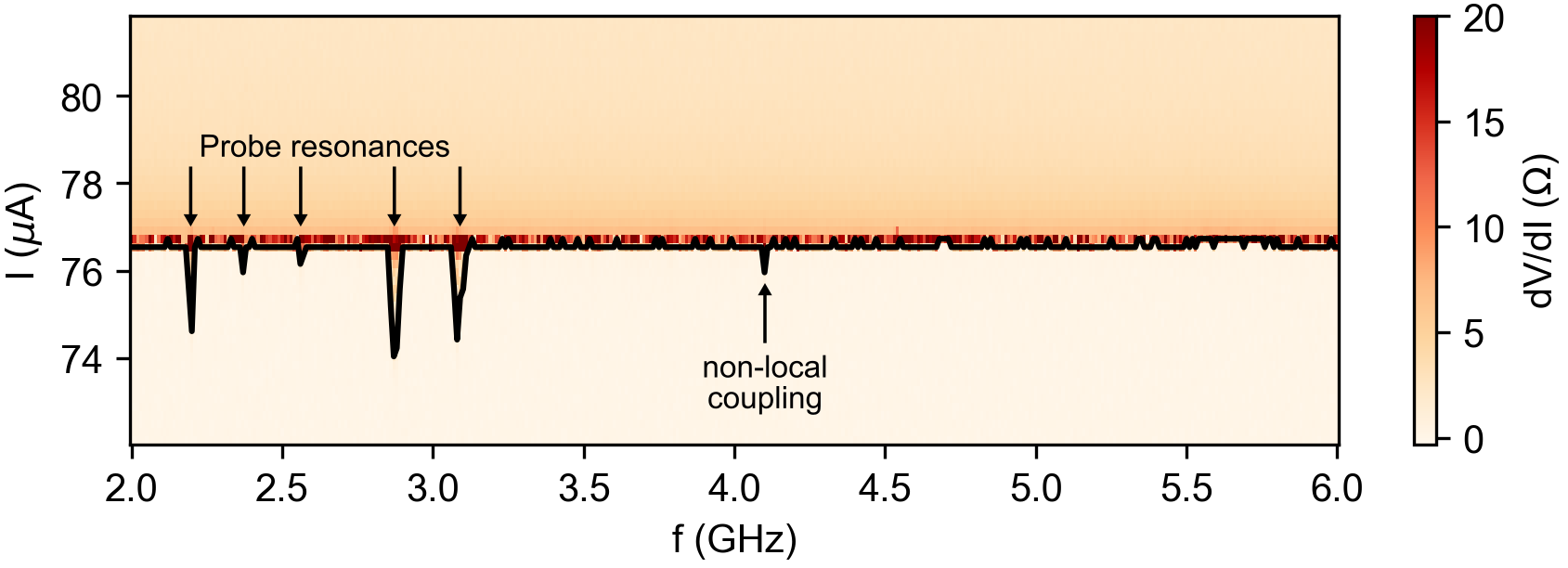}
    \caption{$dV/dI$ as function of DC current $I$ through the SQUID and frequency $f$. The black line indicates the critical current. Several resonances in $I_c$ are observed between 2 and 3.1 GHz, corresponding to self-resonances of the probe. The resonance of the $\lambda/4$ resonator can be seen due to non-local detection by the superconducting contact leads.}
    \label{fig:sup_fig_1}
\end{figure}

First, the microwave environment was characterized using a Krytar 203BK zero-bias Schottky diode detector, which enabled broadband transmission measurements without the use of a vector network analyzer, as shown in Figure \ref{fig:diode}a.
Here, the resonance at 4.102 GHz is barely visible, while the baseline exhibits strong fluctuations. These fluctuations are attributed to impedance mismatch in the flexible cryostat wiring and imperfect shielding. A zoomed-in view of the resonance is shown in Figure \ref{fig:diode}b, where the resonance remains clearly distinguishable despite the pronounced baseline oscillations.
Next, this background is compared with the response of the SQUID when positioned above the central feedline. A signal generator supplies a microwave signal with a power of 0 dBm at frequencies from 2 to 6 GHz, while current-voltage characteristics are measured following the procedure detailed in Section \ref{sec:level2} of the main text. Figure \ref{fig:sup_fig_1} shows the current-voltage characteristics as a function of frequency, where $I_c$ is indicated with a black line. As discussed in the main text, $I_c$ provides a measure for microwave power coupled into the SQUID, such that whenever the measured power is high, $I_c$ will decrease. Between 2 and 3.1 GHz, several dips in $I_c$ appear. As these frequencies are far away from the CPW resonance frequency, these resonances are attributed to self-resonances of the whole probe. This effect is likely associated with an eigenmode of the SQUID tip, leading to particularly efficient absorption at this frequency\cite{davis2019influence}.
They cannot be caused by more power arriving at the sample, as this would be reflected as pronounced peaks in the transmission measurements in Figure \ref{fig:diode}a. There is no correlation between the resonances in Figure \ref{fig:sup_fig_1} and the oscillations in Figure \ref{fig:diode}. Since the resonance at 2.873 GHz shows a high response, this resonance was used to perform the SQUID characterization in section \ref{sec:level2}. It is important to realize that the CPW resonance frequency of the sample  at 4.102 GHz lies far away from the self resonances of the probe, and, that a self resonance is not required to characterize the sample, as shown in Section \ref{sec:level3}. Nevertheless, due to the non-local coupling of the superconducting contact leads, the CPW resonance can be seen in this measurement at 4.102 GHz, even when the tip is not positioned near the resonator.

\subsection{Simulated SQUID response under microwave flux excitation}
\label{sup:even}
\begin{figure}[H]
    \centering
    \includegraphics{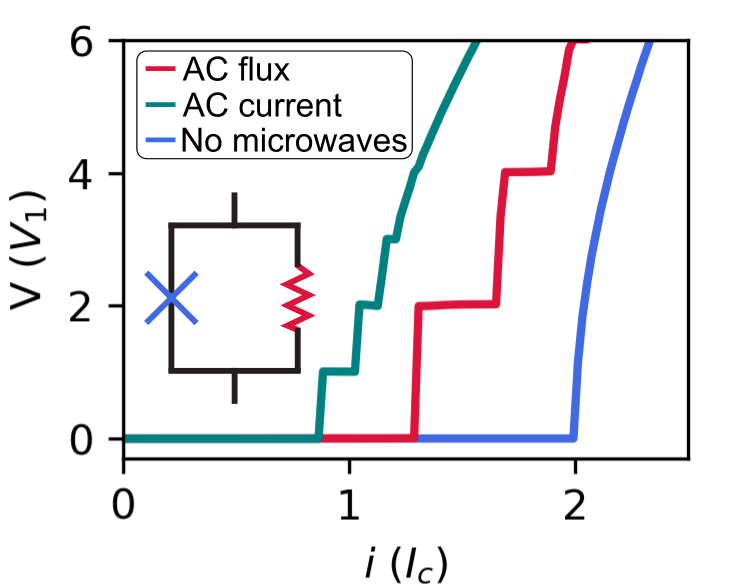}
    \caption{Simulated I-V characteristics of a Josephson junction for three different microwave-driving conditions: no microwave drive (blue), microwave current drive (green), and microwave flux drive (red). The different responses illustrate the distinct Shapiro-step structures produced by current and flux modulation. The inset shows the equivalent resistively shunted junction (RSJ) model used for the simulations.}
    \label{fig:sup_SQUIDsims}
\end{figure}

The resistively-shunted junction (RSJ) model allows to accurately describe single SNS Josephson junctions \cite{panghotra2020giant,Raes2020Shapiro}. The model is schematically shown in the inset of Figure \ref{fig:sup_SQUIDsims}, where the junction can be modeled as a superconducting current channel governed by the Josephson equation
\begin{equation}
    I_s = I_c\sin{\phi},
\end{equation}
with $I_s$ the supercurrent, $I_c$ the critical current and $\phi$ the phase difference -- and a resistive quasiparticle channel with ohmic behavior
\begin{equation}
    I_N = V_j/R_N,
\end{equation}
with $I_N$ the quasiparticle current, $V_j$ the voltage over the junction and $R_N$ the junction resistance, which can be approximated as the constant normal state resistance for SNS junctions. The total current $I$ can be calculated using Kirchoff's current law
\begin{equation}\label{eq:SI_1}
    I = I_c\sin{\Phi} + \frac{V_j}{R_N}.
\end{equation}
Equation \ref{eq:SI_1} can be rewritten in terms of the phase difference using the second Josephson equation ($V_j = \frac{\Phi_0}{2\pi}\frac{d\Phi}{dt}$), leading to
\begin{equation}
    I = I_c\sin{\Phi} + \frac{1}{R_N}\frac{\Phi_0}{2\pi}\frac{d\Phi}{dt}.
\end{equation}
To obtain the data in the right panel of Figure \ref{fig:fig1}d, the RSJ model was solved under a combined DC and microwave bias current: $I= I_{DC} + I_{\mu\text{w}}\cos{2\pi ft}$, where $f$ is the microwave frequency. In the absence of microwave excitation ($I_{\mu\text{w}}=0$), a conventional $IV$ curve without Shapiro steps is replicated. For finite microwave amplitudes ($I_{\mu\text{w}} > 0$), even and odd Shapiro steps emerge at voltages $V_n = n\Phi_0f$. 
To evaluate whether the observed response could originate solely from an oscillating magnetic flux, the microwave current term was set to zero ($I_{\mu\text{w}}=0$) and the critical current was modulated with a time varying flux according to the SQUID interference relation:
\begin{gather}
    I_{c,\text{eff}}=2I_c\cos{\frac{\pi\Phi(t)}{\Phi_0}},\\
    \Phi(t) = \Phi_{DC}+\Phi_{\text{AC}}\sin{2\pi f t},
\end{gather}
with $\Phi_{DC}$ the DC flux offset and $\Phi_{AC}$ the amplitude of the oscillating flux. Numerical solutions of the RSJ equations in this regime produce $IV$ characteristics containing only even Shapiro steps\cite{Mizugaki1999evensteps,Kondo2001evensteps}, as shown in Figure \ref{fig:sup_SQUIDsims}. The observation of both even and odd Shapiro steps in the experimental data (Section \ref{sec:level2}), together with the excellent agreement between experiment and simulations in Figure \ref{fig:fig1}d, demonstrates that the SQUID response cannot be explained by flux modulation alone. Instead, the device is predominantly driven by an oscillating microwave current flowing through the SQUID.

\subsection{SONNET simulations}
\label{sup:sonnet}
\begin{figure}[H]
    \centering
    \includegraphics[width=\linewidth]{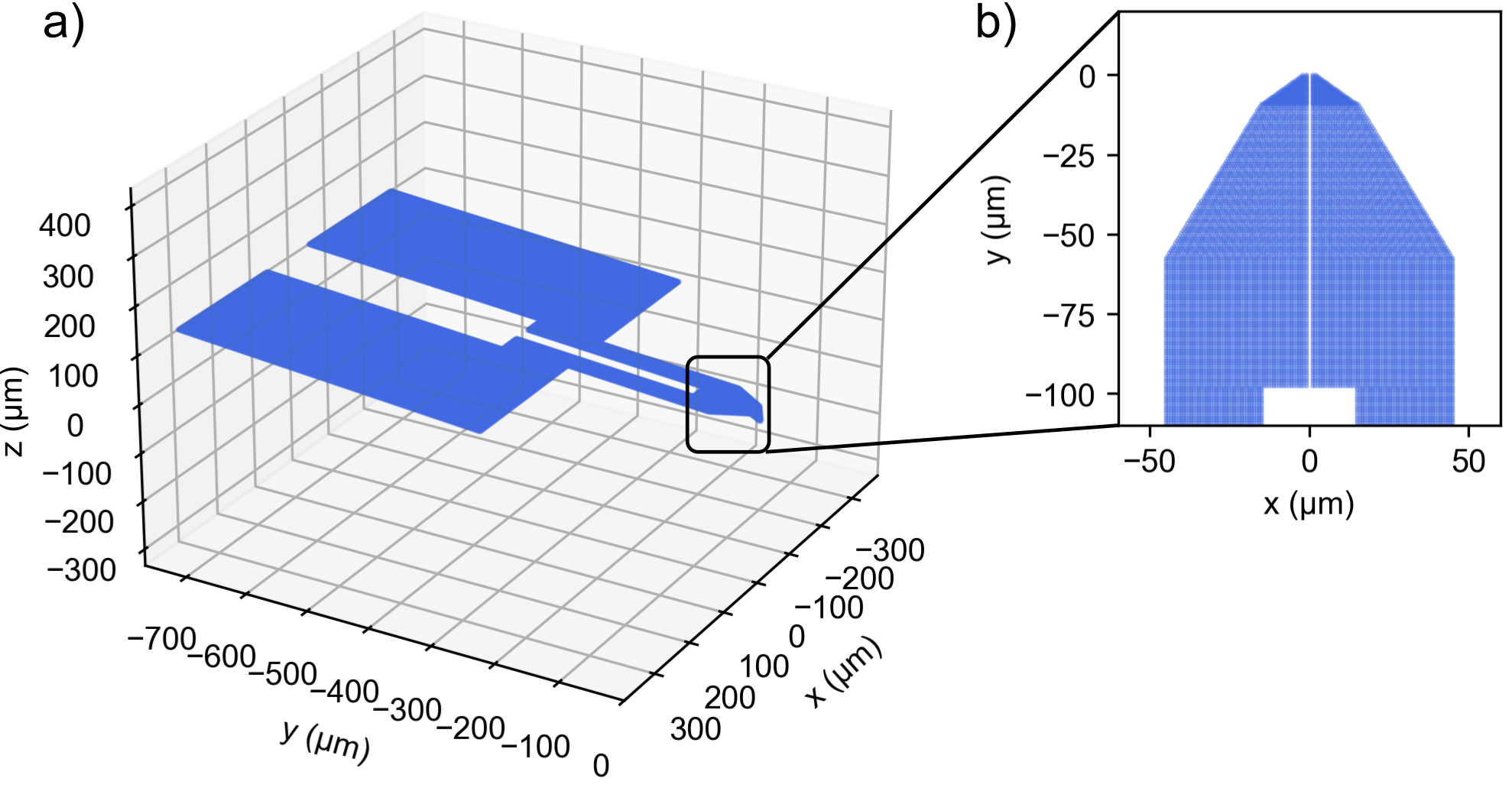}
    \caption{a) Geometrical model of the cantilever used to determine the total flux through the probe. b) Zoom in on the end of the cantilever. }
    \label{fig:sup_sonnet}
\end{figure}

The simulations used in this work are based on the surface current density of the superconducting resonator at the resonance frequency. The current distribution is obtained from Sonnet \cite{Sonnet}, using a method-of-moments electromagnetic solver applied to the exact device geometry. Here, the resonator is modelled as a perfect conductor, resulting in a resonance frequency of 4.8 GHz. From these simulations, Sonnet provides the magnitude and phase of the complex surface current density. This information is used to build up magnetic field maps by numerically solving the Biot-Savart equation via an FFT-based convolution over a range of heights above the device \cite{yazdanian2020biotsavart}. Interpolating these magnetic field maps allows us to construct a 3D volume of the complex magnetic field and consequently estimate the field to be probed at any position in space. The probing cantilever itself is not included in these simulations. Instead, to model the experimental results, a 2D geometric model mimicking the cantilever probe is constructed in this 3D magnetic field map above the resonator, as shown in Figure \ref{fig:sup_sonnet}a. The tip of the probe is maintained at a height of 100 nm above the resonator. However, it is important to note that the cantilever is at an angle with respect to the sample's surface, meaning that the end of the cantilever is higher than the tip of the probe. The furthest parts of the cantilever are over 100 µm from the surface. We do not expect these parts to affect the signal, as the field decays rapidly with height and can be considered negligible there, as demonstrated in SI \ref{sup:height}.
Furthermore, as can be seen in \ref{fig:sup_sonnet}b, we decided to not include the SQUID itself in this probe reconstruction. The resonator, cantilever, and SQUID span vastly different length scales. Resolving the SQUID directly would thus require an impractically fine mesh over the entire probe reconstruction, so we decide to exclude it and instead choose the spacing on the probe surface based on the length scale of the cantilever alone. Finally, the magnetic flux through the probe's surface is calculated at each scan position by numerically integrating the normal component of the magnetic field over the probe area. This allows us to either perform line scans or raster scans, by changing the position of the probe and integrating the magnetic field over its surface at that position.

\subsection{Estimated magnetic field amplitude at the sample}
\label{sup:Bsamp}
Using the procedure described in SI \ref{sup:sonnet}, we can estimate the maximum field amplitude in a perfect-conducting CPW. On the center of the CPW transmission line, at a height of 100 nm, the maximum magnetic field amplitude through the SQUID is approximately 1 $\mu$T for an applied power of $-25$ dBm. From this information and since the magnetic field scales with the square root of microwave power (in Watt), we can estimate the magnetic field at the sample based on how much microwave power is present. From transmission measurements, the cable losses on the input and output line are estimated at $-5$ dB, while the input line has an additional -18 dB attenuation. For the data in Figure \ref{fig:fig1}b, at an applied power of 3 dBm, approximately 3 dBm $- 5$ dBm $- 18$ dBm $= -20$ dBm (10.0 $\mu$W) arrives at the waveguide. Therefore, this power results in a microwave magnetic field amplitude of approximately 1.8 $\mu$T. Table \ref{tbl:conversion} shows an estimate for the magnetic field amplitude through the SQUID for the powers reported in the main text. The table reports the input power from the signal generator, $P_{in}$, the power that arrives at the sample including cable losses, $P_{samp}$, both in units of dBm and $\mu$W, the fraction of power that arrives at the sample compared to the simulation at $-25$ dBm, $P_{samp}/P_{sim}$, the fraction of microwave current that arrives at the sample, $\sqrt{P_{samp}/P_{sim}}$, and the magnetic field at the waveguide $B_{\mu\textrm{w}}$. These values serve as an estimate for the magnetic field at the waveguide. Since the superconducting contact leads couple non-locally, values serve as a lower estimate for the microwave magnetic field at the waveguide.

\begin{table}[H]
  \caption{Conversion from supplied power ($P_{in}$) to estimated maximum microwave magnetic field amplitude ($B_{\mu\textrm{w}}$) at the center of the CPW. The reference power in the simulations, $P_{sim} = -25$ dBm $=3.2$ $\mu$W. }
  \label{tbl:conversion}
  \centering
  \begin{tabular}{lrrrrr}
    \hline
    $P_{in}$ (dBm) & $P_{samp}$ (dBm) & $P_{samp}$ ($\mu$W) & $P_{samp}/P_{sim}$&  $\sqrt{P_{samp}/P_{sim}}$  & $B_{\mu\textrm{w}}$ ($\mu$T) \\
    \hline
    6.0  & $-17.0$ & 20.0  & 6.2 & 2.5  & 2.5  \\
    3.0  & $-20.0$ & 10.0  & 3.1 & 1.8  & 1.8  \\
    2.2  & $-20.8$ & 8.3   & 2.6 & 1.6  & 1.6 \\
    $-8.8$ & $-31.8$ & 0.7   & 0.2 & 0.5  & 0.5\\

    \hline
  \end{tabular}
\end{table}

\subsection{Comparison simulations and experiment at different positions}
\label{sup:bend}
\begin{figure}[H]
    \centering
    \includegraphics{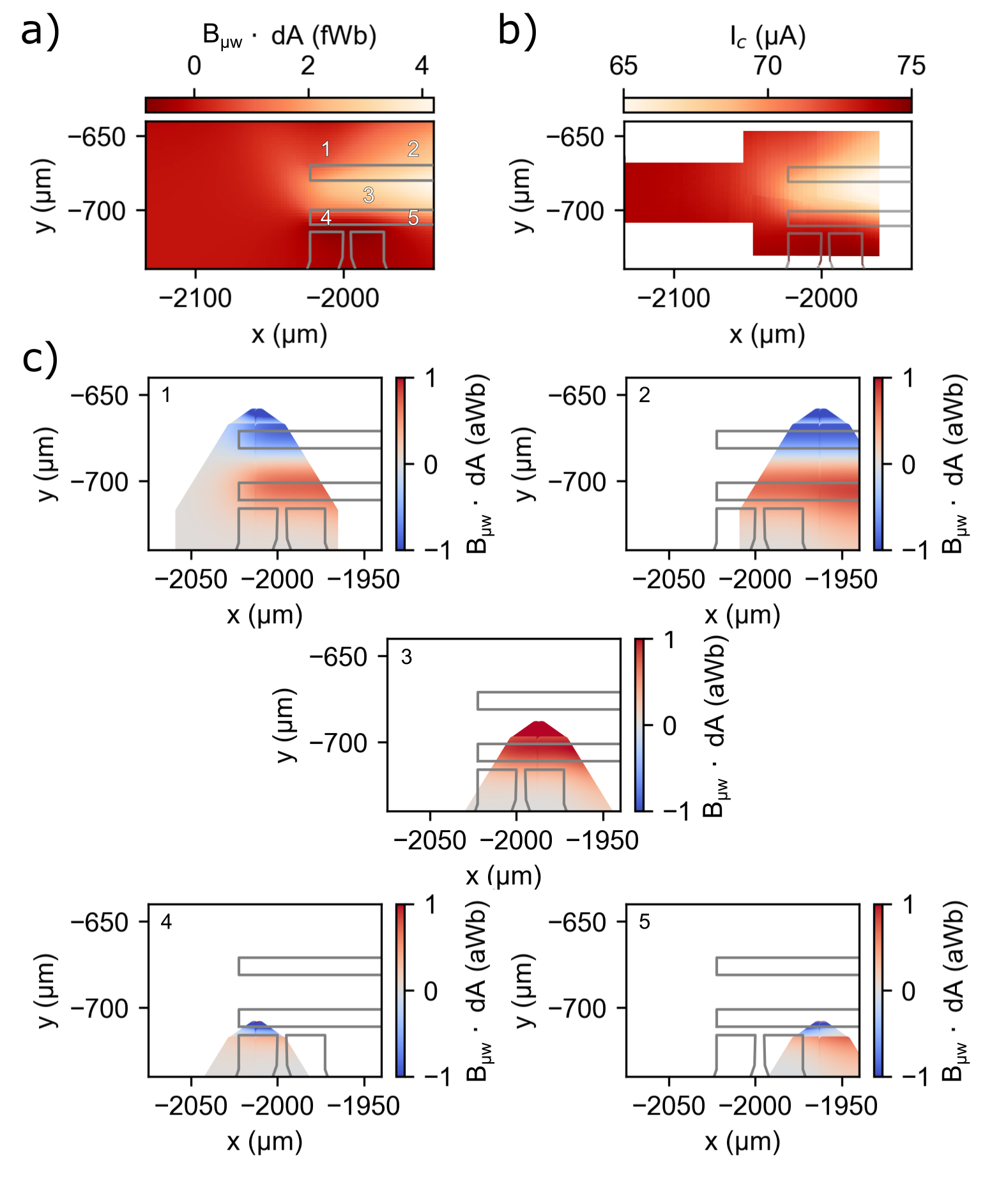}
    \caption{a) Simulated magnetic flux integrated over the complete probe geometry as a function of position near the end of the resonator. b) SQUID $I_c$ as function of position near the end of the resonator. c) Simulated spatial distributions of magnetic flux through the probe geometry for the five positions indicated in panel a}
    \label{fig:supp_end}
\end{figure}

This section further illustrates the influence of the finite spatial extent of the superconducting contacts on the measured response. Figure \ref{fig:supp_end}a shows the simulated magnetic flux integrated over the entire contact geometry as a function of probe position near the end of the resonator, obtained from Sonnet simulations, described in detail in SI \ref{sup:sonnet}. The resonator geometry is indicated by the gray outlines. Figure \ref{fig:supp_end}b extends the analysis presented in Fig. \ref{fig:fig3}b into the surrounding ground plane. The simulations reproduce the experimentally observed behavior, confirming that the modulation of the SQUID critical current originates from microwave flux intercepted by the superconducting contact structure. 
Figure \ref{fig:supp_end}c displays the spatial distribution of magnetic flux through the contact geometry at the five positions indicated in Fig. \ref{fig:supp_end}a. For positions 1 and 2, a large fraction of the contact area overlaps with the resonator, causing the measured signal to reflect the magnetic field generated by the entire resonator section beneath the contacts. In contrast, at positions 3, 4, and 5 only a limited portion of the contacts overlaps with the resonator, resulting in a more localized response. These results demonstrate that the effective spatial resolution is strongly determined by the geometry of the superconducting leads. Consequently, reducing the area of the contacts provides a straightforward route toward improved spatial resolution. 
It should be noted that these simulations only account for the magnetic flux integrated over the contact geometry and are therefore insensitive to local variations and dynamical processes within the superconducting contact leads. In particular, they do not capture interference effects between microwave currents induced in the two contact arms.

\subsection{Height dependence of fields in CPW}
\begin{figure}[H]
    \centering
    \includegraphics{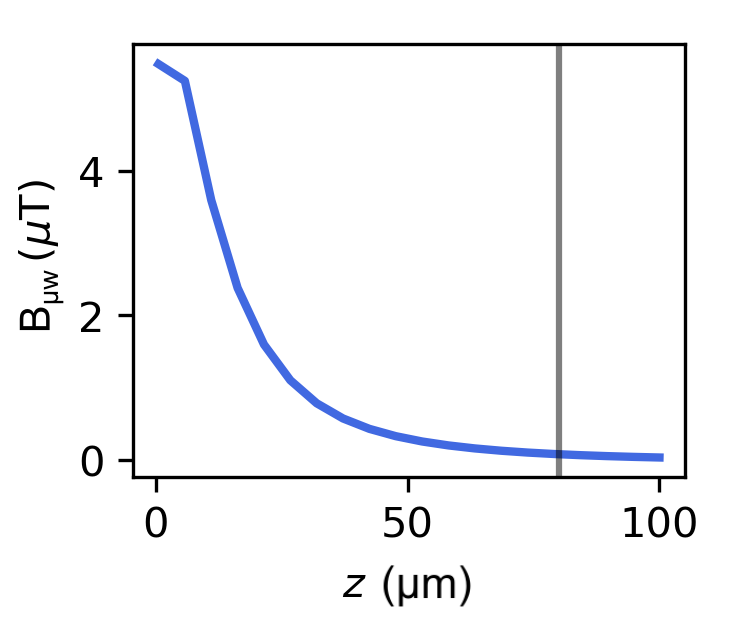}
    \caption{Height dependence of the microwave magnetic field through the SQUID extracted from Sonnet simulations, the gray line indicates 80 $\mu$m above which the field amplitude drops below 1\% of its value at $z=0$.}
    \label{fig:supp_height}
\end{figure}

Figure \ref{fig:supp_height} shows the simulated microwave magnetic field through the SQUID as function of height. The gray line indicates the height above which the field becomes smaller than 1\% of its value at $z=0$.

\label{sup:height}
\bibliography{Ref}

\end{document}